\documentclass[11pt,a4paper]{article}
\usepackage[T1]{fontenc}
\usepackage[utf8]{inputenc}
\usepackage[english]{babel}
\usepackage[margin=25mm,top=23mm,bottom=25mm]{geometry}
\usepackage{mathpazo}
\usepackage{amsmath,amssymb}
\usepackage{microtype}
\usepackage{graphicx}
\usepackage{booktabs,array}
\usepackage{caption}
\usepackage{titlesec}
\usepackage{needspace}
\usepackage{placeins}
\usepackage[authoryear,round]{natbib}
\usepackage{xurl}
\usepackage[hidelinks,unicode]{hyperref}
\usepackage{bookmark}

\hypersetup{
  pdftitle={Reproducible LLM Inference Benchmarking: A Sequential Isolation Protocol for Regression Testing},
  pdfauthor={Arnold Olympio, Juan Manuel Servera Bondroit, Wael Abdelmalek, Guang Lu, João Carvalho},
  pdfkeywords={LLM Inference, Reproducibility, Benchmarking, Tail Latency, Deployment Economics}
}
\titleformat{\section}{\large\bfseries}{\thesection}{0.65em}{}
\titleformat{\subsection}{\normalsize\bfseries}{\thesubsection}{0.65em}{}
\titlespacing*{\section}{0pt}{15pt plus 3pt minus 2pt}{6pt}
\titlespacing*{\subsection}{0pt}{11pt plus 2pt minus 1pt}{4pt}
\renewenvironment{abstract}{\par\noindent\textbf{Abstract}\par\smallskip\noindent\ignorespaces}{\par\vspace{5pt}}
\newlength{\tablewidth}
\newcolumntype{L}[1]{>{\raggedright\arraybackslash}p{#1}}
\newcolumntype{R}[1]{>{\raggedleft\arraybackslash}p{#1}}
\newcolumntype{C}[1]{>{\centering\arraybackslash}p{#1}}

\begin{document}
\begin{center}
{\fontsize{18}{22}\selectfont\bfseries Reproducible LLM Inference Benchmarking:\par A Sequential Isolation Protocol for Regression Testing\par}
\vspace{14pt}
{\normalsize Arnold Olympio\textsuperscript{1}, Juan Manuel Servera Bondroit\textsuperscript{2}, Wael Abdelmalek\textsuperscript{3},\\[3pt] Guang Lu\textsuperscript{1*}, João Carvalho\textsuperscript{3,4}\par}
\vspace{9pt}
{\small
\textsuperscript{1}Lucerne University of Applied Sciences and Arts, Switzerland\\[2pt]
\textsuperscript{2}Microsoft, Switzerland\\[2pt]
\textsuperscript{3}Uthereal AG, Switzerland\\[2pt]
\textsuperscript{4}ETH Zurich, Switzerland\\[2pt]
}
\vspace{3pt}
{\small \textbf{* Correspondence:} Guang Lu\\\href{mailto:guang.lu@hslu.ch}{guang.lu@hslu.ch}\par}
\end{center}
\vspace{3pt}
{\small Keywords: LLM Inference, Reproducibility, Benchmarking, Tail Latency, Deployment Economics.\par}
\vspace{9pt}
\begin{abstract}
Reproducible benchmarking of Large Language Model (LLM) inference is challenging because repeated measurements can vary with execution and system state. We present the Sequential Isolation Methodology, a controlled benchmarking and regression-testing protocol designed to reduce between-run measurement variance while deliberately varying workload concurrency. We evaluate three representative open-source LLMs on an NVIDIA A100 80GB GPU using vLLM 0.9.1 across six context sizes and eight concurrency levels, with five repetitions per configuration. The final protocol reduces average coefficient of variation (CV) from 15.2\% in the least controlled methodology stage to 2.2\% under the final protocol; using CV computed across the five repetition-level median (P50) TTFT values per configuration, 113 of 144 configurations (78.5\%) achieve CV below 3\%. The measurements also show a marked latency transition between 200 and 500 concurrent users on the tested stack and descriptive differences in P99 latency across the three models. We additionally provide an explicit cost break-even model with sensitivity to API pricing. The protocol is intended to provide a stable reference for reproducible comparison and regression testing rather than to predict absolute behavior under uncontrolled production traffic. Infrastructure-as-Code and benchmark scripts support replication of the experimental environment.
\end{abstract}

\hypertarget{introduction}{%
\section{Introduction}\label{introduction}}

The rapid adoption of large language models (LLM) in production systems has created an urgent need for reliable performance benchmarking methodologies. Yet existing benchmarks suffer from a critical flaw: high measurement variance makes results difficult to reproduce and capacity planning unreliable. Published benchmarks often report single-run measurements on heterogeneous hardware without variance quantification, severely limiting their practical utility for deployment decisions (\hyperref[ref:stuhlmann2025]{Stuhlmann et al., 2025}).~

Reproducibility in LLM benchmarking has practical implications for system comparison and regression testing. When repeated benchmark measurements vary substantially, differences between models, configurations or software versions become difficult to distinguish from experimental variation. Our preliminary methodology stage produced an average CV of 15.2\%, motivating a more controlled measurement protocol.

The objective of the present work is not to predict absolute production behavior from an isolated benchmark. Instead, we seek a stable reference measurement against which models, system versions and configurations can be compared fairly. Potential sources of between-run variation include scheduler interactions, residual execution state, initialization effects and sustained device load; however, the reported experiments do not instrument these mechanisms individually. We therefore evaluate whether progressively increasing experimental control reduces observed run-to-run variation rather than claiming a causal decomposition of its sources.

We propose the Sequential Isolation Methodology as a controlled reproducibility and regression-testing protocol. The final benchmark lifecycle executes the test matrix sequentially at the run level, restarts the serving environment when context size changes, applies a standardized warm-up to every run and uses fixed inter-run spacing. The staged V1--V4 evaluation measures the cumulative effect of increasing experimental control; because the design is cumulative rather than factorial, it does not establish the independent causal contribution of each individual control.

This paper makes three contributions. First, we introduce a controlled protocol for reproducible LLM inference benchmarking and regression comparison. Second, we evaluate the protocol using three representative open-source models across 144 model \ensuremath{\times} context-size \ensuremath{\times} concurrency configurations, each repeated five times, giving 720 runs in total. Third, we show how the resulting measurements support descriptive latency comparison, capacity investigation on the tested stack and transparent cost analysis. These results provide controlled reference measurements rather than direct predictions of performance under uncontrolled production traffic.

\hypertarget{related-work}{%
\section{Related Work}\label{related-work}}

While existing work addresses inference optimization and deployment economics independently, no prior study combines reproducible benchmarking methodology with empirical model comparison and cost-performance analysis in a unified framework. Our work directly addresses this gap.

\hypertarget{llm-inference-optimization}{%
\subsection{LLM Inference Optimization}\label{llm-inference-optimization}}

Recent advances in LLM serving focus on memory management and scheduling efficiency. \hyperref[ref:kwon2023]{Kwon et al. (2023)} introduced PagedAttention in vLLM, treating KV cache as virtual memory to achieve 2--4\ensuremath{\times} throughput improvement through reduced memory fragmentation. \hyperref[ref:agrawal2024]{Agrawal et al. (2024)} addressed the throughput-latency tradeoff through Sarathi-Serve's chunked prefill scheduling, demonstrating that naïve continuous batching can degrade tail latency under high load. \hyperref[ref:dao2023]{Dao et al. (2023)} optimized attention computation through FlashAttention-2's improved memory access patterns, achieving significant speedups on modern GPU architectures. The MLPerf Inference benchmark suite (\hyperref[ref:mlcommons2025]{MLCommons, 2025}) has recently expanded to include Llama 3.1 8B with standardized latency targets, though their methodology focuses on throughput rather than variance quantification.

\hypertarget{benchmarking-and-reproducibility}{%
\subsection{Benchmarking and Reproducibility}\label{benchmarking-and-reproducibility}}

Holistic evaluation frameworks like HELM (\hyperref[ref:liang2022]{Liang et al., 2022}) establish best practices for systematic benchmarking, recommending minimum five repetitions with variance reporting to establish statistical confidence. \hyperref[ref:blackwell2024]{Blackwell et al. (2024)} emphasize quantifying uncertainty in LLM benchmark scores, noting that single-run measurements provide insufficient information for production decisions. The LLM-Inference-Bench project (\hyperref[ref:patel2024]{Patel et al., 2024}) provides comprehensive hardware comparisons across AI accelerators but relies primarily on mean latency metrics rather than tail latency analysis. Even established frameworks such as vLLM emphasize the importance of version-pinned dependencies and Infrastructure-as-Code approaches for reproducibility (\hyperref[ref:vllm2025]{vLLM Project, 2025}).

These strands address related but distinct problems. Serving-system research primarily optimizes throughput, latency, scheduling and memory management, while benchmark frameworks standardize what is measured and how systems are compared. The focus of the present work is different: controlling the execution lifecycle of repeated measurements so that model, configuration and software-version comparisons are less affected by between-run variation. Our contribution is therefore a benchmarking methodology layered on an existing serving system, not a new inference architecture or serving optimization.

\hypertarget{deployment-economics}{%
\subsection{Deployment Economics}\label{deployment-economics}}

\hyperref[ref:pan2025]{Pan et al. (2025)} provide a comprehensive cost-benefit framework demonstrating that on-premise deployment reaches break-even at approximately 50 million tokens per month. However, their analysis relies on theoretical throughput estimates rather than empirical measurements under production-realistic loads. \hyperref[ref:chen2023]{Chen et al. (2023)} demonstrate a cost reduction through model cascading, achieving 30-40\% savings while maintaining accuracy thresholds. Our research bridges optimization research and economic analysis by providing reproducible empirical measurements that enable evidence-based deployment decisions combining both performance and cost considerations.

\hypertarget{method}{%
\section{Method}\label{method}}

\hypertarget{the-reproducibility-challenge}{%
\subsection{The Reproducibility Challenge}\label{the-reproducibility-challenge}}

The Sequential Isolation Methodology was developed through four cumulative methodology stages. \hyperref[tab:main1]{Table 1} reports the observed reduction in average CV from 15.2\% in V1 to 2.2\% in V4 as progressively greater experimental control was introduced. Because the stages form a cumulative rather than factorial evaluation, the stage-to-stage changes characterize the protocol as implemented but do not isolate the independent causal contribution of individual controls.

\begin{table}[!htbp]
\centering
\caption{Cumulative Methodology Stages and Observed CV Reduction}
\label{tab:main1}
\small
\setlength{\tablewidth}{\dimexpr\linewidth-6\tabcolsep\relax}
\begin{tabular}{@{}L{0.105\tablewidth}L{0.250\tablewidth}C{0.115\tablewidth}L{0.530\tablewidth}@{}}
\toprule
\textbf{Version} & \textbf{Approach} & \textbf{Avg. CV} & \textbf{Methodology stage} \\
\midrule
V1 & Concurrent testing & 15.2\% & Least controlled baseline \\
V2 & Parallel batches & 10.8\% & Intermediate batching-control stage \\
V3 & Sequential + gaps & 8.3\% & Sequential runs with inter-run spacing \\
V4 & Full isolation & 2.2\% & Final controlled protocol \\
\bottomrule
\end{tabular}
\end{table}

The staged results show progressively lower average CV as the benchmark procedure becomes more controlled. V1 represents the least controlled baseline, V2 introduces an intermediate batching-control stage, V3 uses sequential run execution with inter-run spacing, and V4 applies the final benchmark lifecycle described below. These results establish the cumulative behavior of the protocol; they should not be interpreted as a factorial causal decomposition of scheduler effects, cache state or thermal effects.

\hypertarget{sequential-isolation-methodology}{%
\subsection{\texorpdfstring{Sequential Isolation Methodology }{Sequential Isolation Methodology }}\label{sequential-isolation-methodology}}

``Sequential'' refers to execution of the benchmark matrix and control of the serving lifecycle, not to eliminating concurrency within a test run. For each model, the serving process is started fresh for a context size. All eight target concurrency levels (5, 10, 20, 50, 100, 200, 500 and 1,000 users) are then executed in sequence, with five repetitions at each level, giving 40 runs against the same server instance for that context size. A 10-second interval separates successive runs. The server is terminated after those 40 runs and started fresh for the next context size, reloading the model and resetting resident serving state.

Each run begins with a 30-second warm-up period that is excluded from the measurement window. Concurrent load is generated using Locust with a load-dependent user spawn rate between 2 and 100 users/s. Within a run, each simulated user issues requests sequentially---waiting for its current request to complete before issuing the next---while requests from different users overlap to create the configured concurrency. vLLM uses its default continuous-batching behavior; no custom request-scheduling or batching policy is configured.

The protocol is designed to reduce controllable between-run variability through a standardized execution lifecycle. Scheduler interactions, residual execution state, initialization effects and sustained device load motivate these controls, but GPU clocks, power, scheduler events and prefix-cache statistics were not logged for the reported experiments. We therefore treat these mechanisms qualitatively and do not attribute specific fractions of the observed variance reduction to individual causes.

\hypertarget{experimental-design}{%
\subsection{Experimental Design}\label{experimental-design}}

Our experimental design consists of six context sizes (1k, 2k, 4k, 8k, 16k and 32k tokens), eight concurrent-user levels (5, 10, 20, 50, 100, 200, 500 and 1,000 users), and five repetitions for each model \ensuremath{\times} context-size \ensuremath{\times} concurrency configuration. This gives 48 configurations per model, 240 runs per model and 720 runs across the three models. The context sizes span short prompts through extended-document workloads, while the concurrency levels provide a controlled load range for examining latency behavior on the tested infrastructure.

To replicate typical production workloads such as document summarization, prompts were generated as instruction-following tasks with controlled input lengths matching the target context size (1k--32k tokens) and fixed output generation of 512 tokens. Target concurrency is generated by Locust using a load-dependent spawn rate of 2--100 users/s; each simulated user waits for one request to complete before issuing the next. Three models were selected for this study, i.e. Mistral-7B-Instruct-v0.3 (7.25B parameters, 32k context), Llama-3.1-8B-Instruct (8.03B parameters, 128k context), and Qwen2.5-7B-Instruct (7.6B parameters, 128k context). They were chosen as representative of the 7B-parameter open-source class, covering distinct architectures from Mistral AI, Meta, and Alibaba, all released within the same timeframe (July--September 2024).

All experiments were conducted on version-pinned infrastructure. The hardware configuration used Azure Standard\_NC24ads\_A100\_v4 instances with an NVIDIA A100 80GB GPU, 24 vCPUs and 220 GiB RAM. The software stack included vLLM 0.9.1, CUDA 12.4, PyTorch 2.4.0 and Terraform 1.9.8. The vLLM server used default continuous batching and the flags -\/-gpu-memory-utilization 0.95, -\/-max-model-len 8192 for contexts up to 4k or -\/-max-model-len 32768 for the larger-context runs, -\/-disable-log-requests, and -\/-dtype auto. For the evaluated models, -\/-dtype auto resolves to bfloat16. Benchmark scripts, configurations, analysis code and Infrastructure-as-Code definitions are maintained in the project repository.

\section{Results}\label{results}

\hypertarget{reproducibility-achievement}{%
\subsection{Reproducibility Achievement}\label{reproducibility-achievement}}

\begin{figure}[!htbp]
\centering
\includegraphics[width=\linewidth, trim=0 0 0 10pt, clip]{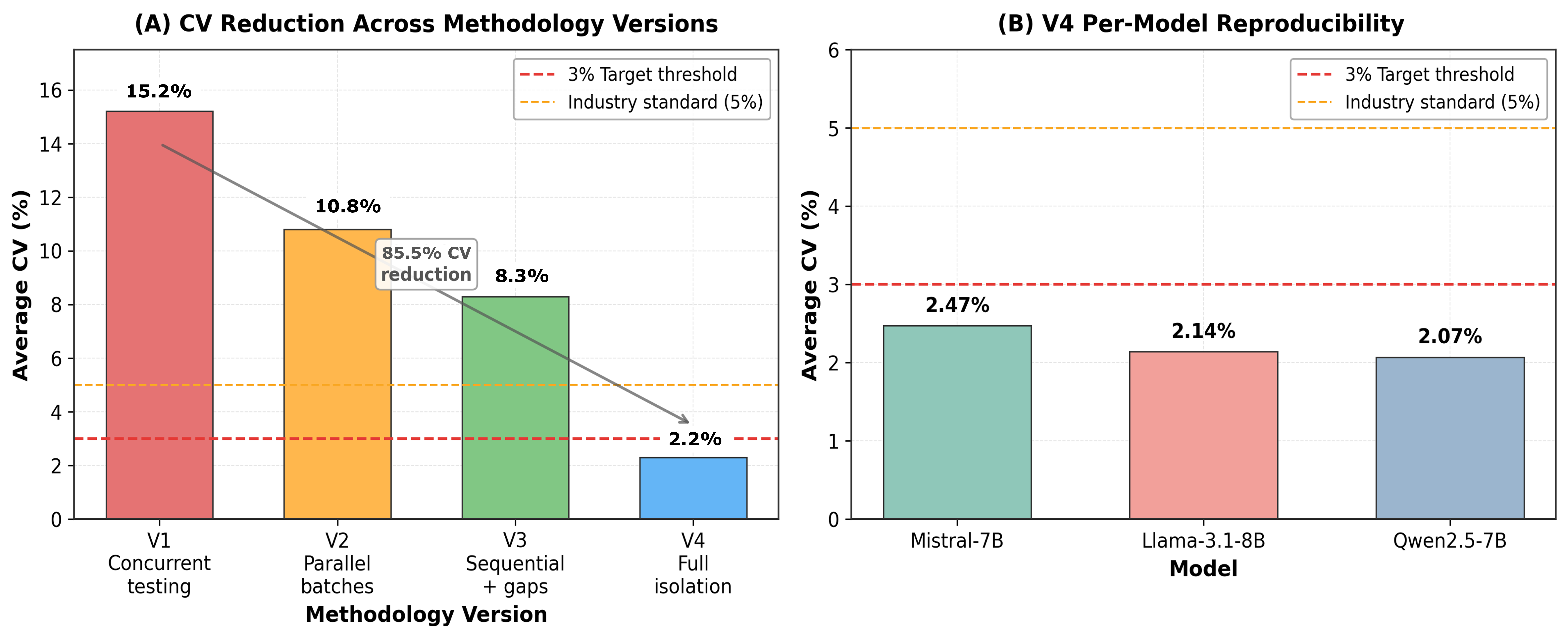}
\caption{Reproducibility across methodology stages and models. Left: average coefficient of variation (CV = \ensuremath{\sigma}/\ensuremath{\mu} \ensuremath{\times} 100\%) of median (P50) TTFT across five repetitions per fixed configuration, summarized for methodology stages V1--V4. Right: average configuration-level CV under the final protocol for each evaluated model. The 3\% line indicates the predefined reference threshold used in the analysis. The 5\% line labelled ``Industry standard'' is shown only as an additional visual reference and is not used in the statistical analysis or conclusions.}
\label{fig:1}
\end{figure}

The final Sequential Isolation protocol achieved an average configuration-level CV of 2.23\% across the evaluated models. CV was computed separately for each fixed model \ensuremath{\times} context-size \ensuremath{\times} concurrency configuration as \ensuremath{\sigma}/\ensuremath{\mu} across the five repetition-level median (P50) TTFT values. Of the 144 configurations, 113 (78.5\%) had CV below the predefined 3\% reference threshold. \hyperref[tab:s1]{Supplementary Table 1} summarizes the configuration-level CV statistics: Qwen2.5-7B had an average CV of 2.07\%, Llama-3.1-8B 2.14\%, and Mistral-7B 2.47\%. \hyperref[fig:1]{Figure 1} summarizes the reduction across methodology stages and the final per-model average CV values.

\hypertarget{queue-saturation-discovery}{%
\subsection{Queue Saturation Discovery}\label{queue-saturation-discovery}}

Because the load-testing matrix does not sample intermediate points between 200 and 500 concurrent users, the precise saturation knee cannot be identified from the present experiments. On the tested A100/vLLM stack, however, the measurements show a marked latency transition within this interval. P50 TTFT rises from 91 ms at 200 users to 12,068 ms at 500 users and 18,542 ms at 1,000 users. We therefore treat 200 users as the highest tested point preceding the observed transition, rather than as an exact or generally applicable production-capacity threshold. Finer-grained experiments between 200 and 500 users would be required to localize the knee more precisely. P50 TTFT is used for this characterization because P99 values at the highest loads approach the request-timeout region, making them less informative for describing further queue growth. \hyperref[tab:s2]{Supplementary Table 2} reports the corresponding values.

The observed transition is consistent with increasing memory pressure and scheduler queueing at high concurrency. However, request-level KV-cache occupancy, scheduler events and related device telemetry were not instrumented in the reported experiments, so the mechanism should be interpreted as a plausible explanation rather than as a directly measured causal result. The observed capacity behavior is specific to the tested hardware, software and workload configuration and should be validated separately under deployment-specific traffic.

\begin{figure}[!htbp]
\centering
\includegraphics[width=\linewidth]{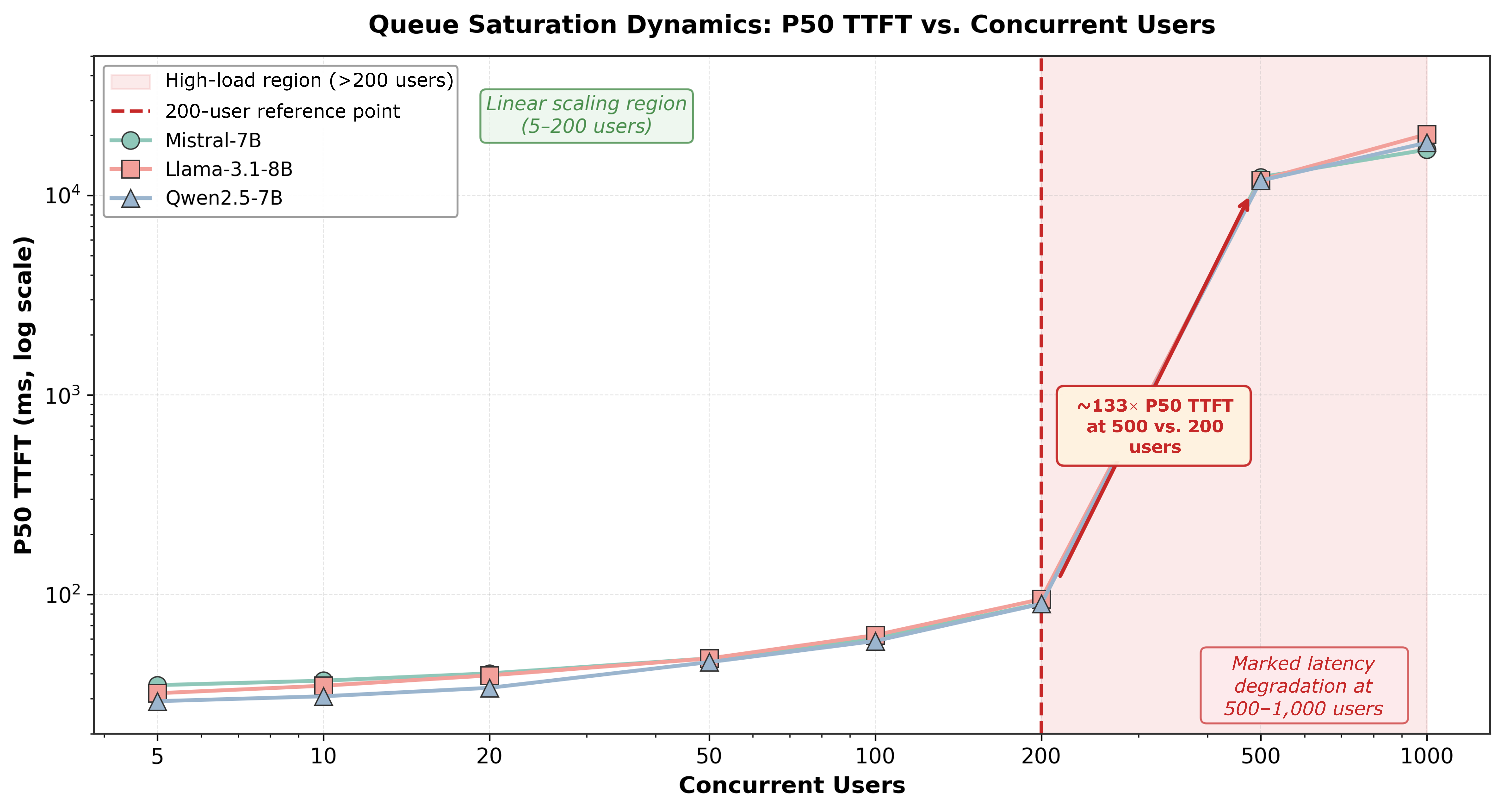}
\caption{Queue saturation dynamics on the tested A100/vLLM stack. P50 TTFT is summarized across the evaluated models and context sizes at each target concurrency level, with five repetitions per fixed configuration. The plot shows the marked latency transition observed between 200 and 500 concurrent users.}
\label{fig:2}
\end{figure}

\hypertarget{three-model-comparison}{%
\subsection{Three-Model Comparison}\label{three-model-comparison}}

The three-model comparison shows descriptive differences in P99 tail latency across the tested configuration matrix. Qwen2.5-7B has the lowest overall average P99 TTFT at 8,022 ms, compared with 8,496 ms for Llama-3.1-8B and 9,124 ms for Mistral-7B, corresponding to observed differences of 5.6\% and 12.1\%, respectively. Given the limited number of repetitions per configuration (n = 5), we report these differences descriptively and do not make inferential significance claims. Average P99 values summarize the eight tested user-load levels (5--1,000 users). \hyperref[tab:s3]{Supplementary Table 3} reports the comparative summaries and \hyperref[tab:s4]{Supplementary Table 4} provides the corresponding per-load results.

\begin{figure}[!htbp]
\centering
\includegraphics[width=\linewidth]{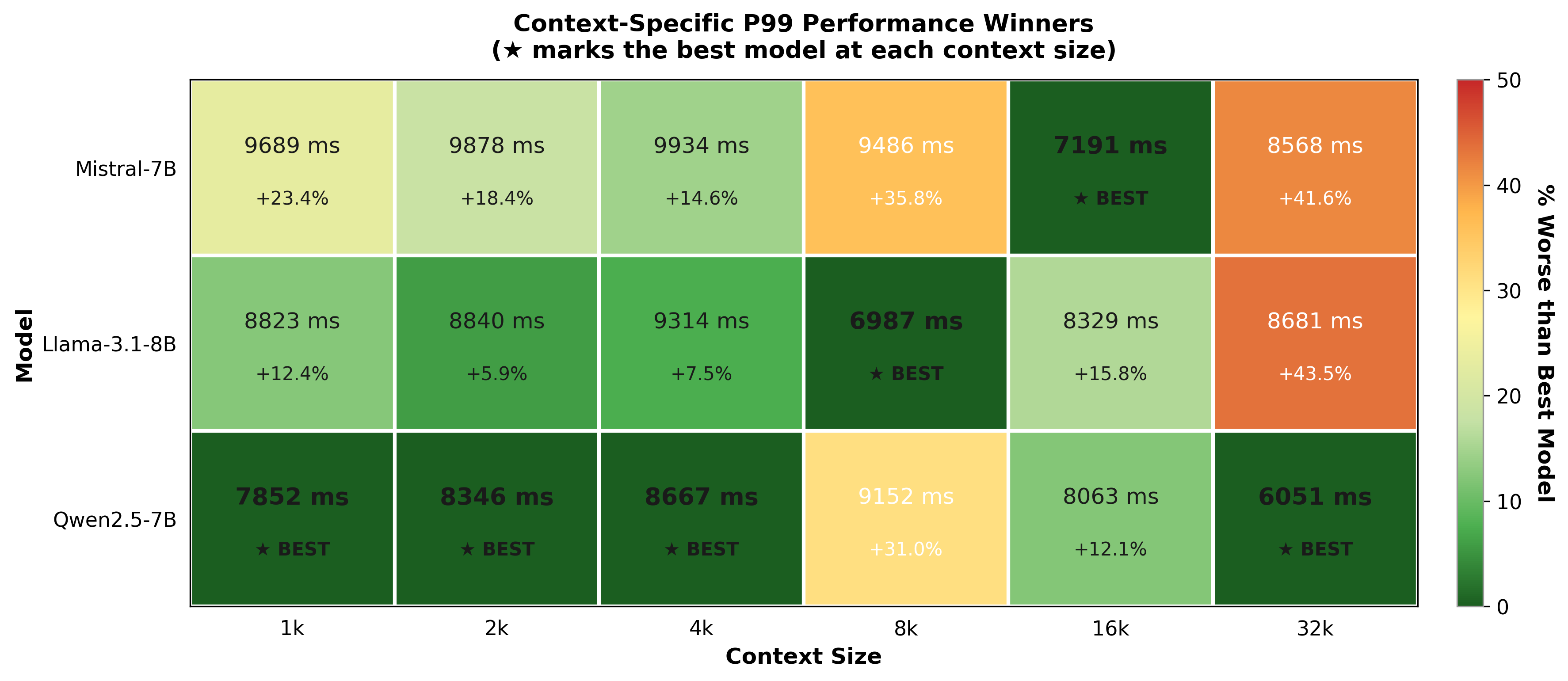}
\caption{Context-specific P99 TTFT comparison under the final protocol. Values summarize the tested concurrency levels for each model and context size, with five repetitions per fixed configuration. The lowest observed P99 TTFT at each context size is highlighted descriptively.}
\label{fig:3}
\end{figure}

Context-specific analysis reveals nuanced performance patterns that inform deployment strategies. Qwen2.5-7B has the lowest observed P99 latency at short contexts (1k--4k tokens) with 12.8--18.9\% P99 advantages and excels at the longest tested context (32k tokens) with a 29.4\% advantage over Mistral. However, Llama-3.1-8B has the lowest observed P99 latency at 8k context length with a 23.7\% P99 advantage, while Mistral-7B achieves the best results at 16k context with a 10.8\% advantage. \hyperref[tab:s5]{Supplementary Table 5} summarizes these context-specific winning configurations. Beyond latency, inter-token latency ranges from 13.8 ms at minimal load to 66.5 ms at maximum load, corresponding to generation rates of 15--72 tokens per second across all three models. \hyperref[fig:3]{Figure 3} summarizes context-specific P99 winners as a heatmap.

\hypertarget{the-32k-context-bimodal-anomaly}{%
\subsection{The 32k Context Bimodal Anomaly}\label{the-32k-context-bimodal-anomaly}}

All three models show a pronounced gap between median and tail latency at 32k context length, with P50 values of approximately 90--750 ms and P99 values of 6,051--8,681 ms. The resulting P50-to-P99 multipliers differ substantially across models: approximately 61\ensuremath{\times} for Mistral-7B, 95\ensuremath{\times} for Llama-3.1-8B and 8\ensuremath{\times} for Qwen2.5-7B. This pattern indicates substantially greater tail-latency variability at the longest tested context. Differences in memory and cache state are plausible contributing mechanisms, but cache hit/miss behavior was not instrumented directly. The result should therefore be interpreted descriptively rather than as evidence of a specific cache mechanism.

\hypertarget{throughput-analysis}{%
\subsection{Throughput Analysis}\label{throughput-analysis}}

\hyperref[tab:s6]{Supplementary Table 6} reports per-request throughput in tokens per second (tok/s), derived from inter-token latency (ITL) P50 across all context sizes at each stable load level (\ensuremath{\leq}200 concurrent users). At minimal load (5 users), Qwen2.5-7B achieves the highest throughput at 70.2 tok/s, followed by Mistral-7B at 67.8 tok/s and Llama-3.1-8B at 63.6 tok/s. Under peak stable load (200 users), Mistral-7B demonstrates superior throughput retention at 26.5 tok/s versus Qwen2.5-7B (21.9 tok/s) and Llama-3.1-8B (20.6 tok/s), reflecting a 61\% degradation for Mistral versus 69\% for both Qwen and Llama. Averaged across all stable loads, Mistral-7B leads at 48.5 tok/s, Qwen2.5-7B at 47.9 tok/s, and Llama-3.1-8B at 42.6 tok/s. These results suggest that for throughput-sensitive deployments, Qwen excels at low concurrency while Mistral maintains more stable output rates as concurrent load increases.

\hypertarget{discussion}{%
\section{Discussion}\label{discussion}}

\hypertarget{why-the-sequential-isolation-methodology-achieves-reproducibility}{%
\subsection{Why the Sequential Isolation Methodology Achieves Reproducibility}\label{why-the-sequential-isolation-methodology-achieves-reproducibility}}

The average CV decreases from 15.2\% in V1 to 2.2\% in V4 as the benchmark procedure becomes progressively more controlled. The individual phenomena motivating the controls---scheduler interactions under concurrent load, state carry-over between runs, warm-up effects and sustained device load---are known qualitatively and are not claimed as novel here. The reported experiments did not log GPU clocks, power, scheduler events or prefix-cache statistics at sufficient granularity to establish a causal decomposition of the variance reduction. The contribution is therefore the controlled benchmarking protocol and its staged evaluation rather than a new causal account of these mechanisms. Because the V1--V4 design is cumulative rather than factorial, the independent effect of each individual control cannot be separated from the present data.

\hypertarget{implications-for-benchmarking-and-deployment-evaluation}{%
\subsection{Implications for Benchmarking and Deployment Evaluation}\label{implications-for-benchmarking-and-deployment-evaluation}}

The measurements are most appropriately interpreted as reproducible reference values for relative comparison on the tested stack. They may inform deployment hypotheses, but they should not be treated as direct predictions of absolute behavior under uncontrolled production traffic. On the evaluated A100/vLLM setup, latency rises sharply between 200 and 500 concurrent users, and the relative model ordering varies with context length. Such results can support regression testing, configuration comparison and selection of candidate settings for subsequent production-like validation. Representative traffic traces, arrival processes and deployment-specific state behavior should be evaluated separately before operational conclusions are drawn.

\hypertarget{cost-performance-trade-off-analysis}{%
\subsection{Cost-Performance Trade-off Analysis}\label{cost-performance-trade-off-analysis}}

Using Azure Standard\_NC24ads\_A100\_v4 pricing of \$3.07 per hour and 730 billed hours per month, a continuously provisioned instance costs approximately \$2,241 per month. Break-even against API-based deployment is given by \(V^{*} = (c \times h)/p\), where \(c\) is GPU cost per hour, \(h\) is billed hours per month and \(p\) is API cost per token. Because API prices vary, we report a sensitivity range rather than a single break-even figure. At \$0.60 per million tokens, break-even is approximately 3.7 billion tokens per month (approximately 124 million tokens/day); at \$0.15 per million tokens, it rises to approximately 14.9 billion tokens per month (approximately 498 million tokens/day). Measured single-instance throughput at 70\% utilization is approximately 288 million tokens per month, well below this range. Under these assumptions, token-price economics alone therefore do not justify continuously provisioned self-hosting on a single GPU; aggregate scale and non-price considerations such as latency control, customization and data residency may change the decision. Provider pricing, utilization, engineering overhead and other operating costs vary by deployment, so these values should be interpreted as an illustrative sensitivity analysis rather than a universal threshold.

This economic analysis complements \hyperref[ref:pan2025]{Pan et al.'s (2025)} deployment framework by relating the cost comparison to empirically measured single-instance throughput. Under the assumptions used here, the calculated break-even range lies well above the measured capacity of a single instance, indicating that token-price economics alone would require aggregate deployment scale beyond one GPU. Non-price considerations such as latency control, customization and data residency may nevertheless influence deployment decisions.

\hypertarget{conclusion}{%
\section{Conclusion}\label{conclusion}}

We presented the Sequential Isolation Methodology as a controlled protocol for reproducible LLM inference benchmarking and regression comparison. Across 144 model \ensuremath{\times} context-size \ensuremath{\times} concurrency configurations, each repeated five times, the final protocol achieved an average CV of 2.23\%, with 113 configurations (78.5\%) below the predefined 3\% reference threshold. On the tested A100/vLLM stack, we observed a marked latency transition between 200 and 500 concurrent users and descriptive differences in P99 latency across the three models. The cost analysis further shows that, under the stated assumptions, API/self-hosting break-even lies between approximately 3.7 and 14.9 billion tokens per month depending on API pricing. These findings provide controlled reference measurements; they should not be interpreted as predictions of absolute performance under uncontrolled production traffic.

Future work will validate the protocol across additional serving frameworks, including SGLang and TensorRT-LLM; newer GPU generations such as H100 and H200; and newer dense and Mixture-of-Experts model families. We will also evaluate representative and alternative synthetic workloads, including ShareGPT/LMSYS-style traces, Poisson arrival processes and long-context workloads, and independently vary output-length, arrival-rate and cache-state regimes. A further priority is comparison of controlled reference measurements with concurrent production-like serving to determine when the former do and do not transfer to operational workloads. Finally, a factorial ablation with direct system telemetry would be needed to separate the contribution of individual controls, while finer-grained sampling between 200 and 500 concurrent users would allow more precise localization of the observed latency transition.

\hypertarget{limitation}{%
\section{Limitation}\label{limitation}}

Several limitations constrain the external validity of the findings. First, all experiments were conducted on a single NVIDIA A100 80GB GPU with vLLM 0.9.1; behavior may differ across newer GPU generations, serving frameworks and model architectures. Second, the study uses controlled synthetic prompts with a fixed 512-token output target. Alternative output-length distributions and representative production traces were not evaluated, and the Locust arrival pattern was not independently varied as an experimental factor. Third, the server is restarted when context size changes rather than before every individual concurrency-level repetition, so serving state is not independently reset for every run. Fourth, the protocol provides a controlled reference measurement and has not been validated as a predictor of concurrent production behavior. Fifth, the V1--V4 evaluation is cumulative rather than factorial, so the independent causal effects of individual controls cannot be separated. Finally, GPU clocks, power, scheduler events and cache statistics were not logged for the reported experiments, limiting causal interpretation of the mechanisms underlying the observed variance reduction.

\hypertarget{conflict-of-interest}{%
\section{Conflict of Interest}\label{conflict-of-interest}}

The authors declare that the research was conducted in the absence of any commercial or financial relationships that could be construed as a potential conflict of interest.

\hypertarget{author-contributions}{%
\section{Author Contributions}\label{author-contributions}}

AO: Conceptualization, Methodology, Software, Investigation, Data curation, Formal analysis, Visualization, Writing -- original draft. JC: Conceptualization, Supervision, Project administration, Writing -- review \& editing. GL: Supervision, Writing -- review \& editing. WA: Conceptualization, Writing -- review \& editing. JMSB: Conceptualization, Writing -- review \& editing. All authors approved the submitted version.

\hypertarget{funding}{%
\section{Funding}\label{funding}}

This research received no external funding.

\hypertarget{acknowledgments}{%
\section{Acknowledgments}\label{acknowledgments}}

The authors thank UTHEREAL.ai for proposing the research direction and providing the industry context that motivated this work, and the Lucerne University of Applied Sciences and Arts (HSLU) for supporting this research. This study extends work originally conducted as part of the first author\textquotesingle s Master\textquotesingle s thesis at HSLU.

\hypertarget{generative-ai-statement}{%
\section{Generative AI Statement}\label{generative-ai-statement}}

The authors used Claude (Anthropic) and ChatGPT (OpenAI) to assist with language editing, restructuring text, improving clarity and concision, and revising the manuscript. The tools were not used to generate experimental data or independently determine scientific results or conclusions. All AI-assisted text and revisions were reviewed and verified against the underlying experimental data and benchmark scripts by the authors, who take full responsibility for the content.


\hypertarget{data-availability-statement}{%
\section{Data Availability Statement}\label{data-availability-statement}}

The benchmark outputs, benchmark scripts, configurations, analysis code and Infrastructure-as-Code (Terraform) definitions underlying this study are maintained in the project GitHub repository. The repository is currently private. We plan to make a sanitized version of the code and data publicly available, with credentials and repository-state files removed.

\clearpage

\setcounter{section}{0}

\setcounter{table}{0}

\renewcommand{\theHsection}{supp.\arabic{section}}

\renewcommand{\theHtable}{supp.\arabic{table}}

\renewcommand{\tablename}{Supplementary Table}

\section*{Supplementary Material}

\addcontentsline{toc}{section}{Supplementary Material}

\hypertarget{supplementary-tables}{%
\section{Supplementary Tables}\label{supplementary-tables}}

\textbf{Supplementary Note S1. Benchmark execution lifecycle}

For each model, the benchmark server is started fresh for each context size. The eight target concurrency levels (5, 10, 20, 50, 100, 200, 500 and 1,000 users) are executed sequentially, with five repetitions at each level, giving 40 runs against the same serving instance for each context size. A 10-second interval separates successive runs. Each run begins with a 30-second warm-up phase that is excluded from measurement. Concurrent users are generated with Locust using a load-dependent spawn rate between 2 and 100 users/s; each individual simulated user waits for its current request to complete before issuing the next request. The server is terminated and restarted when the context size changes. vLLM uses default continuous batching with no custom scheduling policy. Across three models, six context sizes and eight concurrency levels, the experiment contains 144 configurations and 720 repeated runs.

\begin{table}[!htbp]
\centering
\caption{Reproducibility Results Across Models (Configuration-Level CV Statistics). Coefficient of Variation (CV = \ensuremath{\sigma}/\ensuremath{\mu} \ensuremath{\times} 100\%) is computed separately for each fixed model \ensuremath{\times} context-size \ensuremath{\times} concurrency configuration across its five repetition-level median (P50) TTFT values under the final Sequential Isolation protocol. Experiments use vLLM 0.9.1 on an Azure A100 80GB GPU with -\/-dtype auto (resolving to bfloat16 for the evaluated models). There are 48 configurations per model and 144 configurations overall. Across all models, 113 of 144 configurations (78.5\%) have CV below the predefined 3\% reference threshold.}
\label{tab:s1}
\small
\setlength{\tablewidth}{\dimexpr\linewidth-6\tabcolsep\relax}
\begin{tabular}{@{}L{0.340\tablewidth}R{0.220\tablewidth}R{0.220\tablewidth}R{0.220\tablewidth}@{}}
\toprule
\textbf{Model} & \textbf{Avg CV} & \textbf{Best CV} & \textbf{Worst CV} \\
\midrule
Mistral-7B & 2.47\% & 0.17\% & 13.28\% \\
Llama-3.1-8B & 2.14\% & 0.36\% & 5.96\% \\
Qwen2.5-7B & 2.07\% & 0.16\% & 6.89\% \\
Overall & 2.23\% & 0.16\% & 13.28\% \\
\bottomrule
\end{tabular}
\end{table}

\begin{table}[!htbp]
\centering
\caption{Queue Saturation Dynamics. P50 TTFT values are summarized across the three evaluated models and six context sizes at each concurrent-user level on the tested Azure A100 80GB/vLLM 0.9.1 stack. The degradation factor is computed relative to the 200-user measurement, the highest tested load preceding the marked latency transition observed between 200 and 500 users. The mechanism underlying this transition was not instrumented directly; increasing memory pressure and scheduler queueing are plausible contributors.}
\label{tab:s2}
\small
\setlength{\tablewidth}{\dimexpr\linewidth-4\tabcolsep\relax}
\begin{tabular}{@{}C{0.290\tablewidth}R{0.340\tablewidth}L{0.370\tablewidth}@{}}
\toprule
\textbf{Concurrent Users} & \textbf{Avg P50 TTFT (ms)} & \textbf{Degradation Factor} \\
\midrule
200 & 91 & baseline (stable) \\
500 & 12,068 & 133\ensuremath{\times} (saturated) \\
1000 & 18,542 & 204\ensuremath{\times} (saturated) \\
\bottomrule
\end{tabular}
\end{table}

\begin{table}[!htbp]
\centering
\caption{Descriptive P99 Tail-Latency Comparison Across Models. Average P99 TTFT is summarized across 48 configurations per model (6 context sizes \ensuremath{\times} 8 user loads), with five repetitions per configuration, giving 240 runs per model. The ``\% vs Mistral'' column reports the observed relative difference using Mistral-7B as the reference. The ``Load Wins'' column reports the number of the eight tested concurrency levels at which each model has the lowest observed P99 TTFT. These comparisons are descriptive; no inferential significance claim is made. Full per-load results are provided in \hyperref[tab:s4]{Supplementary Table 4}.}
\label{tab:s3}
\small
\setlength{\tablewidth}{\dimexpr\linewidth-6\tabcolsep\relax}
\begin{tabular}{@{}L{0.330\tablewidth}R{0.240\tablewidth}R{0.230\tablewidth}C{0.200\tablewidth}@{}}
\toprule
\textbf{Model} & \textbf{Avg P99 (ms)} & \textbf{vs Mistral} & \textbf{Load Wins} \\
\midrule
Qwen2.5-7B & 8,022 & -12.1\% & 5/8 \\
Llama-3.1-8B & 8,496 & -6.9\% & 1/8 \\
Mistral-7B & 9,124 & baseline & 2/8 \\
\bottomrule
\end{tabular}
\end{table}

\begin{table}[!htbp]
\centering
\caption{Complete P99 TTFT Results by User Load (All Models). P99 Time to First Token (TTFT) values averaged across all six context sizes (1k--32k tokens) for each user load level, computed across the five repetitions per configuration under the V4 Sequential Isolation Methodology. Loads \ensuremath{\leq}200 users represent the stable operating regime; loads \ensuremath{\geq}500 users fall in the saturated regime, where P99 values approach vLLM's per-request timeout and should be interpreted as characterizing system failure behavior rather than normal operation.}
\label{tab:s4}
\small
\setlength{\tablewidth}{\dimexpr\linewidth-8\tabcolsep\relax}
\begin{tabular}{@{}C{0.120\tablewidth}R{0.230\tablewidth}R{0.230\tablewidth}R{0.230\tablewidth}L{0.190\tablewidth}@{}}
\toprule
\textbf{Users} & \textbf{Mistral P99} & \textbf{Llama P99} & \textbf{Qwen P99} & \textbf{Winner} \\
\midrule
5 & 314ms & 249ms & 238ms & Qwen \\
10 & 93ms & 82ms & 118ms & Llama \\
20 & 107ms & 147ms & 111ms & Mistral \\
50 & 150ms & 148ms & 140ms & Qwen \\
100 & 268ms & 272ms & 263ms & Qwen \\
200 & 386ms & 372ms & 330ms & Qwen \\
500 & 23,490ms & 27,702ms & 23,919ms & Mistral \\
1000 & 48,187ms & 38,994ms & 38,555ms & Qwen \\
\bottomrule
\end{tabular}
\par\medskip
\begin{minipage}{\linewidth}\small
\emph{Note: Winner determined by lowest P99 TTFT at each user load level. Qwen wins 5/8 load levels (5, 50, 100, 200, 1000 users), Mistral wins 2/8 (20, 500 users), and Llama wins 1/8 (10 users). At 500 users, Mistral (23,490ms) outperforms Qwen (23,919ms) by 429ms (1.8\%).}
\end{minipage}
\end{table}

\begin{table}[!htbp]
\centering
\caption{Context-Specific Lowest Observed P99 and Performance Differences. Model with the lowest observed P99 TTFT at each context size based on P99 TTFT averaged across all eight user load levels. The ``Advantage'' column reports the percentage improvement of the model with the lowest observed P99 TTFT over the next-best model (for single-context entries) or the range of improvements across the 1k--4k contexts. This summary provides a descriptive comparison of context-specific model performance on the tested setup.}
\label{tab:s5}
\small
\setlength{\tablewidth}{\dimexpr\linewidth-6\tabcolsep\relax}
\begin{tabular}{@{}L{0.220\tablewidth}L{0.320\tablewidth}R{0.240\tablewidth}R{0.220\tablewidth}@{}}
\toprule
\textbf{Context Size} & \textbf{Lowest observed model} & \textbf{P99 (ms)} & \textbf{Advantage} \\
\midrule
1k--4k & Qwen & 7,852--8,667 & 12.8--18.9\% \\
8k & Llama & 6,987 & 23.7\% \\
16k & Mistral & 7,191 & 10.8\% \\
32k & Qwen & 6,051 & 29.4\% \\
\bottomrule
\end{tabular}
\end{table}

\begin{table}[!htbp]
\centering
\caption{Per-Request Throughput (tok/s) by Concurrent User Load --- Average Across Context Sizes (\ensuremath{\leq}200 Users). Throughput values are derived from P50 inter-token latency (ITL) as 1000/ITL, then averaged across the six context sizes (1k--32k tokens) at each user load. Only stable operating loads (\ensuremath{\leq}200 concurrent users) are reported; saturated loads (\ensuremath{\geq}500 users) are excluded because ITL becomes dominated by queueing delay rather than generation speed. All measurements use the final Sequential Isolation protocol on an Azure A100 80GB GPU with vLLM 0.9.1 and -\/-dtype auto, which resolves to bfloat16 for the evaluated models.}
\label{tab:s6}
\small
\setlength{\tablewidth}{\dimexpr\linewidth-6\tabcolsep\relax}
\begin{tabular}{@{}C{0.220\tablewidth}R{0.260\tablewidth}R{0.260\tablewidth}R{0.260\tablewidth}@{}}
\toprule
\textbf{Concurrent Users} & \textbf{Mistral-7B (tok/s)} & \textbf{Llama-3.1-8B (tok/s)} & \textbf{Qwen2.5-7B (tok/s)} \\
\midrule
5 & 67.8 & 63.6 & 70.2 \\
10 & 61.6 & 55.5 & 64.0 \\
20 & 52.0 & 46.6 & 55.7 \\
50 & 45.2 & 38.5 & 42.1 \\
100 & 38.0 & 31.1 & 33.8 \\
200 & 26.5 & 20.6 & 21.9 \\
\bottomrule
\end{tabular}
\end{table}

\begin{table}[!htbp]
\centering
\caption{Descriptive Model Comparison by Workload Requirement. This table summarizes the lowest observed latency or throughput result for selected conditions on the tested single-A100/vLLM setup. The comparisons are specific to the present benchmark and should not be interpreted as general production recommendations; different hardware, serving frameworks and workloads may change the relative ordering.}
\label{tab:s7}
\small
\setlength{\tablewidth}{\dimexpr\linewidth-4\tabcolsep\relax}
\begin{tabular}{@{}L{0.310\tablewidth}L{0.240\tablewidth}L{0.450\tablewidth}@{}}
\toprule
\textbf{Requirement} & \textbf{Best observed model} & \textbf{Rationale} \\
\midrule
Strict P99 SLA & Qwen2.5-7B & Best tail latency control (12.1\% advantage) \\
8k context workloads & Llama-3.1-8B & Context-specific optimization (23.7\% P99 advantage) \\
Higher throughput across stable tested loads & Mistral-7B & Highest average per-request throughput across \ensuremath{\leq}200-user loads (48.5 tok/s) \\
\bottomrule
\end{tabular}
\end{table}

\begin{table}[!htbp]
\centering
\caption{Cost Sensitivity and Break-Even Analysis. A continuously provisioned Azure Standard\_NC24ads\_A100\_v4 instance at \$3.07/hour costs approximately \$2,241/month using 730 billed hours. Break-even volume is calculated as \(V^{*} = (c \times h)/p\), where \(p\) is the API price per token. Because API pricing varies, the table reports sensitivity across illustrative prices of \$0.15--\$0.60 per million tokens. Measured single-instance throughput at 70\% utilization is approximately 288 million tokens/month, below the full break-even range. Engineering, storage, networking and other operating overheads are not included, so the comparison is indicative rather than universal.}
\label{tab:s8}
\small
\setlength{\tablewidth}{\dimexpr\linewidth-4\tabcolsep\relax}
\begin{tabular}{@{}L{0.300\tablewidth}R{0.380\tablewidth}R{0.320\tablewidth}@{}}
\toprule
\textbf{API price} & \textbf{Break-even tokens/month} & \textbf{Approx. Tokens/day} \\
\midrule
\$0.15 / million & 14.9 billion & 498 million \\
\$0.30 / million & 7.5 billion & 249 million \\
\$0.60 / million & 3.7 billion & 124 million \\
\bottomrule
\end{tabular}
\end{table}


\begin{thebibliography}{99}
\bibitem[Agrawal et~al.(2024)]{agrawal2024}\phantomsection\label{ref:agrawal2024}
Agrawal, A., Kedia, N., Panwar, A., Mohan, J., Kwatra, N., Gulavani, B. S., et al. (2024). Taming Throughput-Latency Tradeoff in LLM Inference with Sarathi-Serve. In: Proceedings of the 18th USENIX Symposium on Operating Systems Design and Implementation (OSDI'24).

\bibitem[Blackwell et~al.(2024)]{blackwell2024}\phantomsection\label{ref:blackwell2024}
Blackwell, R. E., Barry, J., \& Cohn, A. G. (2024). Towards Reproducible LLM Evaluation: Quantifying Uncertainty in LLM Benchmark Scores. arXiv preprint arXiv:2410.03492.

\bibitem[Chen et~al.(2023)]{chen2023}\phantomsection\label{ref:chen2023}
Chen, L., Zaharia, M., \& Zou, J. (2023). FrugalGPT: How to Use Large Language Models While Reducing Cost and Improving Performance. arXiv preprint arXiv:2305.05176.

\bibitem[Dao et~al.(2023)]{dao2023}\phantomsection\label{ref:dao2023}
Dao, T., Fu, D. Y., Ermon, S., Rudra, A., \& Ré, C. (2023). FlashAttention-2: Faster Attention with Better Parallelism and Work Partitioning. arXiv preprint arXiv:2307.08691.

\bibitem[Jiang et~al.(2023)]{jiang2023}\phantomsection\label{ref:jiang2023}
Jiang, A. Q., Sablayrolles, A., Mensch, A., Bamford, C., Chaplot, D. S., de las Casas, D., et al. (2023). Mistral 7B. arXiv preprint arXiv:2310.06825.

\bibitem[Kwon et~al.(2023)]{kwon2023}\phantomsection\label{ref:kwon2023}
Kwon, W., Li, Z., Zhuang, S., Sheng, Y., Zheng, L., Yu, C. H., et al. (2023). Efficient Memory Management for Large Language Model Serving with PagedAttention. In Proceedings of the 29th ACM Symposium on Operating Systems Principles (SOSP'23).

\bibitem[Liang et~al.(2022)]{liang2022}\phantomsection\label{ref:liang2022}
Liang, P., Bommasani, R., Lee, T., Tsipras, D., Soylu, D., Yasunaga, M., et al. (2022). Holistic Evaluation of Language Models. arXiv preprint arXiv:2211.09110.

\bibitem[Li et~al.(2024)]{li2024}\phantomsection\label{ref:li2024}
Li, B., Jiang, Y., Gadepally, V., \& Tiwari, D. (2024). LLM Inference Serving: Survey of Recent Advances and Opportunities. arXiv preprint arXiv:2407.12391.

\bibitem[Llama Team(2024)]{llama2024}\phantomsection\label{ref:llama2024}
Llama Team. (2024). The Llama 3 Herd of Models. arXiv preprint arXiv:2407.21783.

\bibitem[MLCommons(2025)]{mlcommons2025}\phantomsection\label{ref:mlcommons2025}
MLCommons. (2025). MLPerf Inference v5.1 Benchmark Results. Retrieved from \url{https://mlcommons.org/2025/09/mlperf-inference-v5-1-results/}.

\bibitem[Pan et~al.(2025)]{pan2025}\phantomsection\label{ref:pan2025}
Pan, G., Chodnekar, V., Roy, A., \& Wang, H. (2025). A Cost-Benefit Analysis of On-Premises Large Language Model Deployment: Breaking Even with Commercial LLM Services. arXiv preprint arXiv:2509.18101.

\bibitem[Patel et~al.(2024)]{patel2024}\phantomsection\label{ref:patel2024}
Patel, P., Choukse, E., Zhang, C., Shah, A., Goiri, I., Maleki, S., et al. (2024). LLM-Inference-Bench: Inference Benchmarking of Large Language Models on AI Accelerators. arXiv preprint arXiv:2411.00136.

\bibitem[Qwen Team(2024)]{qwen2024}\phantomsection\label{ref:qwen2024}
Qwen Team. (2024). Qwen2.5 Technical Report. arXiv preprint arXiv:2412.15115.

\bibitem[Stuhlmann(2025)]{stuhlmann2025}\phantomsection\label{ref:stuhlmann2025}
Stuhlmann, L. (2025). Bench360: Benchmarking Local LLM Inference from 360 Degrees. arXiv preprint arXiv:2511.16682.

\bibitem[vLLM Project(2025)]{vllm2025}\phantomsection\label{ref:vllm2025}
vLLM Project. (2025). vLLM 2024 Retrospective and 2025 Vision. Retrieved from \url{https://blog.vllm.ai/2025/01/10/vllm-2024-wrapped-2025-vision.html}.

\bibitem[Zhong et~al.(2024)]{zhong2024}\phantomsection\label{ref:zhong2024}
Zhong, Y., Liu, S., Chen, J., Hu, J., Zhu, Y., Liu, X., et al. (2024). DistServe: Disaggregating Prefill and Decoding for Goodput-optimized Large Language Model Serving. In: Proceedings of the 18th USENIX Symposium on Operating Systems Design and Implementation (OSDI'24).

\end{thebibliography}
\end{document}